\documentclass{aip-cp}

\usepackage[numbers]{natbib}
\usepackage{rotating}
\usepackage{graphicx}
\usepackage{color}
\usepackage{thmtools}
\declaretheorem{theorem}
\declaretheorem{lemma}

\newcommand{\CHAIN}[1]{\mathbf{#1}}
\renewcommand{\P}{{\CHAIN{P}}}
\newcommand{\Q}{{\CHAIN{Q}}}

\begin{document}

\title{A Classification of Event Sequences in the Influence Network}

\author[aff1]{James Lyons Walsh\corref{cor1}}
\author[aff1,aff2]{Kevin H. Knuth}
\eaddress{kknuth@albany.edu}

\affil[aff1]{Department of Physics, University at Albany (SUNY), 1400 Washington Avenue, Albany, NY, USA, 12222}
\affil[aff2]{Department of Informatics, University at Albany (SUNY)}
\corresp[cor1]{Corresponding author: jlwalsh@albany.edu}

\maketitle

\begin{abstract}
We build on the classification in \cite{Walsh+Knuth:MaxEnt2015} of event sequences in the influence network as respecting \textit{collinearity} or not, so as to determine in future work what phenomena arise in each case. Collinearity enables each observer to uniquely associate each particle event of influencing with one of the observer's own events, even in the case of events of influencing the other observer. We further classify events as to whether they are \textit{spacetime}  events that obey in the fine-grained case the coarse-grained conditions of \cite{Knuth+Bahreyni:JMP2014}, finding that Newton's First and Second Laws of motion are obeyed at spacetime events. A proof of Newton's Third Law under particular circumstances is also presented.
\end{abstract}

\section{Background}

Information physics considers physical laws to result from the consistent quantification of information one possesses about physical phenomena \cite{Knuth:infophysics}.  In previous efforts, we have shown that a simple model of particles that directly influence one another results in a partially ordered set (poset) referred to as the influence network \cite{Knuth:Info-Based:2014}. Both particles and observers are represented by totally ordered chains of influence events.  Consistent quantification of information about events of a particle influencing a pair of coordinated observers results in a unique description of the particle chain's events characterized by the Minkowski metric and Lorentz transformations of special relativity \cite{Knuth+Bahreyni:JMP2014}. When a particle, the events of which are collinear with their projections onto exactly two observers, is itself influenced, the observers describe it as following trajectories in space and time given by equations of the form of geodesic equations from general relativity in 1+1 dimensions \cite{Walsh+Knuth:MaxEnt2015}. That is, a particle in this influence network is described by the mathematics that describes motion through space and time.

The influence network consists of events, each of which represents an act of influencing or being influenced, and the influences, each of which is a relation between one event of one type and one event of the other. The asymmetry between the two types of events introduces a partial order. Within the poset are totally ordered subsets called \textit{chains}. Since the events composing chains are totally ordered, they can be associated with valuations that are integers or real numbers, so that the order in the poset, as far as the observers can quantify it, can be expressed with the greater-than-or-equal ordering of numbers in the valuation \cite{Knuth+Bahreyni:JMP2014}. 

The chains considered here are either particle chains or \textit{observer} chains. For brevity, these will be referred to here as ``particles'' and ``observers.'' Observers are distinguished chains, the valuations of the events of which can be associated with some of the poset events not on the observer chains through the process of \textit{projection}. \textit{Forward projection} of an event to a chain is the following of influence transitively from the event to the least event on the chain that \textit{includes}, or is greater in the order than, the projected event. Similarly, \textit{back projection} is the following of influence transitively from an event to the greatest event on the chain that it includes. Through projection to observer chains, some events in the poset can be assigned sets of numbers constituting the valuations of the events to which they project on the observer chains, giving rise to a coordinate system \cite{Knuth+Bahreyni:JMP2014}.

The most convenient observer chains are \textit{coordinated} ones, which form a set of observers between the events of which there exists a bijectivity and that agree on lengths, which are simply differences in valuations between the endpoints of an interval, the agreement existing in the sense that the endpoints of an interval on one member of a pair of coordinated observers forward and back project to events on the other observer that are the endpoints of intervals on that observer of the same length as the interval on the first observer. The convenience provided by coordinated observers arises from the fact that events can be completely characterized in the coordinate system using only forward projections to the observers, corresponding to influences received by the observers  \cite{Knuth+Bahreyni:JMP2014}. All observers referred to herein are assumed to be  coordinated.

A pair of coordinated observers defines a 1+1 dimensional subspace of the influence network through the concept of \textit{proper collinearity}. An event is \textit{collinear} with its projections onto two chains iff both its projections onto one chain can be found by first projecting onto the other chain and further projecting thence onto the one chain. Collinearity is \textit{proper} iff all projections are invariant under reversing the ordering relation \cite{Knuth+Bahreyni:JMP2014}.

The endpoints of an interval on a particle chain forward project to the endpoints of an interval on one observer, which we call $\P$, of length $dp$ and on the other observer, called $\Q$, of length $dq$. A change of variables to
\begin{equation}\label{dt_and_dx}
dt = \frac{dp + dq}{2} \qquad \qquad \mathrm{and} \qquad \qquad dx = \frac{dp - dq}{2}
\end{equation}
gives the Minkowski metric, $d s^2 = dt^2 - dx^2$, in which $ds$ is the invariant interval length, $d s = \sqrt{dpdq}$, and gives an analogue to velocity, $v = \frac{dx}{dt}$ \cite{Knuth+Bahreyni:JMP2014}. To relate it to other quantities, we name the length of the interval along $\P$ between successive events of being influenced by a particle as $k_R k$. The Lorentz transformations derived in \cite{Knuth+Bahreyni:JMP2014} show that $k = 1$ for observers that measure the particle's velocity as zero and give 
a second expression for velocity, which is equivalent to the first under conditions described below, namely $v = \frac{k - 1/k}{k + 1/k}$. The length of the interval along $\Q$ between successive events of being influenced by the particle is found to be $\frac{k_R}{k}$. In the following, $k_R$ will be set equal to 1, a condition we call \textit{network units}. Cases of other values of $k_R$ will be explored in future work.

The only points in the influence network are events, so that intervals can only be defined as ending at events. \textit{Atomic intervals} are intervals on a particle chain that cannot be subdivided into smaller intervals of nonzero length, the number of atomic intervals in a particle chain interval being called $N$. Because particle interval lengths are defined in terms of the coordinate system determined by the observers, so that a particle chain interval, the endpoints of which project to the same events on both observers, has zero length, $N$ is also one less than the number of events on the observer chains to which events on the closed particle chain interval project. Thus, rates at which the particle chain events project to the observers can be defined \cite{Knuth:Info-Based:2014}, $r_p = \frac{N}{dp}$ being the rate for observer $\P$ and $r_q = \frac{N}{dq}$ being the rate for $\Q$. With another change of variables \cite{Knuth:Info-Based:2014},
\begin{equation}\label{eq:E_and_P}
E = \frac{r_q + r_p}{2} \qquad \qquad \mathrm{and} \qquad \qquad P = \frac{r_q - r_p}{2},
\end{equation}
an analogue to the energy-momentum relation, $E^2 - P^2 = M^2$, is found, where mass is identified as the geometric mean of the rates: $M = \sqrt{r_q r_p}$ \cite{Knuth:Info-Based:2014}. This mass is the rate corresponding to the interval length $ds = \sqrt{dpdq}$. A future publication will discuss the rate connected via $k_R$ to the proper time, which is defined below.

In this paper, we examine certain influence sequences to determine whether they obey collinearity. We expect that future work may show that noncollinear events give rise to phenomena particular to them. We also name as \textit{spacetime} and explore a property of events introduced in \cite{Walsh+Knuth:MaxEnt2015}, finding that it gives rise to Newton's Laws of motion.

\section{Influence Sequences Classified as Not Collinear}

\textbf{Definition 1} \textit{Distinctness (of events)}: Two events are called \textit{distinct} if they 
can be distinguished by the observers so that they cannot be considered a single event.

Sufficient conditions for distinctness are the following.

1) The events lie on different chains that can be independently shown not to intersect at the events.

2) The events have different positions in the order on the same chain.

3) The events project to any chain at events that can be independently established to be distinct.

4) The other members of the events' influencing-influenced pairs can be independently established to be distinct.

5) The events are of different types, influencing or being influenced.

To the observers, events that are not distinct are not distinguishable and thus can only be described as the same event.

\begin{figure}
\centering
\includegraphics[width=0.75\linewidth]{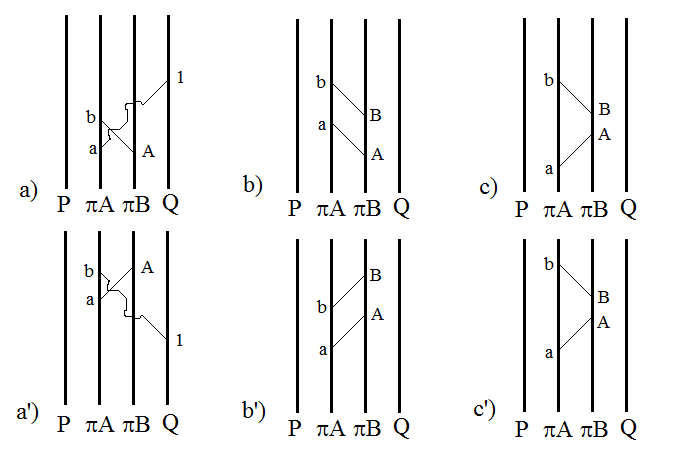}
\caption{These are the event sequences considered in Theorem \ref{thm:new_thm_2_sequence_classification}. Chains $\pi_A$ and $\pi_B$ are particles and $\P$ and $\Q$ observers. 
Some events associated with the observers influencing the other chains are not pictured. Each Hasse diagram denoted by a primed letter in the lower row is the diagram denoted by an unprimed letter immediately above it with the ordering relation reversed, which corresponds to 
reflection through a horizontal line through the center of the diagram, so that top and bottom are exchanged.}
\label{fig:Sequence_Classification}
\end{figure}

\begin{theorem}\label{thm:new_thm_2_sequence_classification}
The particle chain events in Figures \ref{fig:Sequence_Classification} cannot obey proper collinearity.
\end{theorem}

\textbf{Proof} All events in Figure \ref{fig:Sequence_Classification} must be distinct, since otherwise, the referents of two different labels would be quantified as the same event.

Figure \ref{fig:Sequence_Classification}a shows a situation that was shown in \cite{Walsh+Knuth:MaxEnt2015} to violate collinearity.

In Figure \ref{fig:Sequence_Classification}b, since there are no events between a and b or A and B, collinearity would require events a and b to forward project and back project to any other chain at the same events, making them occupy the same position in the order. The same would be true of events A and B. Thus, since there would be no basis for distinctness of a and b or of A and B in the collinear case, this situation violates collinearity.

In Figure \ref{fig:Sequence_Classification}c, we can show the violation of collinearity by contradiction. Collinearity with its projections onto $\P$ and particle chain $\pi_B$ requires that event a can be back projected to $\pi_B$ by forward projecting to $\P$ and back projecting via event b to B. Thus, B would be included by a. However, B includes A, which includes a, so that by transitivity \cite{Knuth:Info-Based:2014}, B includes a. So long as events a and B occupy different positions in the order, as is implied by the fact that B appears higher in the diagram than a, this is a contradiction showing  that collinearity is violated by this event sequence.

Each of the primed figures in the lower row of Figure \ref{fig:Sequence_Classification} is the unprimed figure immediately above it reflected through a horizontal line so that the top of the figure is exchanged with the bottom. That is, the lower row is the upper row with the ordering relation reversed. In the properly collinear case, reversal of the ordering relation leaves the projections unchanged. Thus, any diagram that is properly collinear would remain so under the reversal. You can also reverse the conditional and say that Figure \ref{fig:Sequence_Classification}a is properly collinear if Figure \ref{fig:Sequence_Classification}a$'$ is. Since a is not, a$'$ is not. The same reasoning also shows that Figures \ref{fig:Sequence_Classification}b$'$ and c$'$ are not properly collinear.

\section{Spacetime Events and Additional Properties of the Influence Network}

\textbf{Definition 2} \textit{Spacetime event, spacetime interval, atomic pure spacetime interval, pure spacetime interval, atomic spacetime interval}: A \textit{spacetime event} is an event properly collinear with its projections onto the observers that is an endpoint of a \textit{spacetime interval}, an interval the length of which can be written as a function $ds(dp,dq)$ solely of the total lengths $dp$ and $dq$ along the observer chains \cite{Knuth+Bahreyni:JMP2014}.  An \textit{atomic pure spacetime interval} is a spacetime interval that does not contain a spacetime point, except at its boundaries. A \textit{pure spacetime interval} is a spacetime interval consisting of the union of contiguous atomic pure spacetime intervals and no other intervals. An \textit{atomic spacetime interval} is a spacetime interval that cannot contain spacetime subintervals that have intersection of nonzero length, other than those that are part of a single pure spacetime interval. 

That is, spacetime events are those bounding intervals obeying the conditions  corresponding to the coarse-grained case in \cite{Knuth+Bahreyni:JMP2014}. From these conditions and the more fundamental one that intervals along chains join by set union, so that associativity holds, Knuth and Bahreyni derived the following results that will be used here.

0) For an atomic spacetime interval $a = [\pi_i,\pi_j]$ followed by a pure atomic spacetime interval $b = [\pi_j,\pi_k]$, where the intervals share exactly one event, it holds that $ds_c = ds_a + ds_b$, where $c = [\pi_i,\pi_k]$. Here, we must restrict the statement to intervals of the stated types because the result of \cite{Walsh+Knuth:MaxEnt2015} shows that the condition does not hold for spacetime intervals generally.

1) The interval length on a spacetime interval computed from counting atomic particle chain intervals equals the interval length measured by the observers: $\frac{N}{2} = ds = \sqrt{dpdq}$.

2) The velocity measured by the observers on a spacetime interval equals the velocity arising from the Lorentz transformations: $\frac{dx}{dt} = \frac{dp - dq}{dp + dq} = \frac{k - 1/k}{k + 1/k}$.

Note that not all intervals bounded by spacetime events need obey the conditions, but rather it is the case that all spacetime events bound some interval obeying the conditions. This distinction is important, for example, in considering the situation explored in  \cite{Walsh+Knuth:MaxEnt2015}, which involved reception of influence immediately following a pure atomic spacetime interval giving rise to a non-spacetime interval bounded by spacetime events, so that the union of the intervals was an atomic spacetime interval. 

An item to be addressed in the future is the possibility that particle chain events  may be assigned valuations other than half-integers, corresponding to $k_R \neq 1$,  which would alter condition 1) by a multiplicative factor. As noted above, we work in network units, or set $k_R = 1$, which makes the length calculated as the difference in valuations of events along a particle chain having $N$ atomic intervals equal to $\frac{N}{2}$. We call the length along a particle chain found in this way the \textit{proper time} $d \tau$. Therefore, condition 1) above for a spacetime event is the equality of $d \tau$ with the same length as measured by the observers, $d \tau = ds = \sqrt{dpdq}$. 

\begin{theorem}\label{thm:old_thm_2_localization}
Only at its spacetime events can a particle be localized at a spacetime point, meaning that the observers can assign the particle a value of position $x$ and of
 time $t$ \cite{Walsh+Knuth:MaxEnt2015}.
\end{theorem}

\textbf{Proof} In \cite{Knuth+Bahreyni:JMP2014}, Knuth and Bahreyni showed that associativity of combining intervals, proper collinearity, and the ability to write the interval length along a chain as a function $ds$ of only the corresponding total lengths $dp$ and $dq$ along coordinated observer chains $\P$ and $\Q$, gave rise to the conditions 0) through 2) following Definition 2. They then showed that the change of variables to $dt$ and $dx$ given in equation (\ref{dt_and_dx}) yielded the Minkowski metric, justifying identification of $t$ and $x$ with time and space. When conditions 0)-2) are not obeyed, the identification does not hold. For example, when $d \tau = \frac{N}{2} \neq ds =  \sqrt{dpdq}$, the interval length $ds$ appearing in $ds^2 = dt^2 - dx^2$ in \cite{Knuth+Bahreyni:JMP2014} is not the interval length along the particle chain $d \tau$, so that the Minkowski metric is not obeyed, and $t$ and $x$ cannot be identified as time and space. Hence, location in spacetime is defined only at spacetime events.

\begin{theorem}\label{thm:old_thm_3_spacetime}
On a spacetime interval containing $N$ atomic intervals, the rates at which the particle influences as measured by observer $\P$, $r_p = \frac{N}{dp}$, and as measured by $\Q$, $r_q = \frac{N}{dq}$ \cite{Knuth:Info-Based:2014}, must equal $\frac{2}{k}$ and $2k$, respectively in network units, or equivalently, $dp = \frac{N}{2} k$ and $dq = \frac{N}{2k}$ over the spacetime interval ending in the spacetime event.
\end{theorem}

\textbf{Proof} By defining $A \dot{=} \frac{dp}{k}$ and $B \dot{=} k dq$, condition 2 of Definition 2 can be written as
\begin{equation}
\frac{Ak - \frac{B}{k}}{Ak + \frac{B}{k}} = \frac{k - \frac{1}{k}}{k + \frac{1}{k}}.
\end{equation}
A little algebra gives $A = B$. Substituting the definitions of $A$ and $B$ into condition 1 of Definition 2 then gives $A = \frac{N}{2}$. This shows that $dp = \frac{N}{2} k$ and $dq = \frac{N}{2k}$. Further substituting into the definitions of the rates gives $r_p = \frac{2}{k}$ and $r_q = 2k$.

\begin{lemma}\label{thm:old_lemma_3-1_mass_is_2}
The mass of a particle as measured by the observers is 2 in network units at spacetime events. 
\end{lemma}

\textbf{Proof} By the definition of \cite{Knuth:Info-Based:2014}, mass is $M = \sqrt{r_p r_q} = \sqrt{\frac{2}{k} 2k} = 2$.

As a check on these results, consider a portion of the derivation from \cite{Walsh+Knuth:MaxEnt2015}. It was shown in \cite{Walsh+Knuth:MaxEnt2015} that if a particle received influence at a spacetime event from another particle on the $\Q$ side of the first, the previous event on the receiving particle's chain was constrained to be a spacetime event of influencing $\P$. With unprimed quantities referring to the spacetime interval ending in the first spacetime event and primed to the spacetime interval ending in the second, this gives $N' = N + 1$ and $dq' = dq$, there being no advance along $\Q$ associated with an act of influencing $\P$, since the successive events involved would project to the same event on $\Q$. Since Theorem \ref{thm:old_thm_3_spacetime} gives $dq = \frac{N}{2k}$ and $dq' = \frac{N+1}{2k'}$, this gives $k' = \frac{N+1}{N} k$, as in \cite{Walsh+Knuth:MaxEnt2015}. The expression for $dq'$ in the previous sentence and the fact that $dp' = \frac{N+1}{2} k'$, both required by Theorem \ref{thm:old_thm_3_spacetime}, complete the result in \cite{Walsh+Knuth:MaxEnt2015}. Note that since we are considering values of $k$ that change and defining the coefficient of $k$ in $dp$ and of $\frac{1}{k}$ in $dq$ so that in $dp = N_p k$ and $dq = N_q \frac{1}{k}$ \cite{Walsh+Knuth:MaxEnt2015}, $N_p$ and $N_q$ are no longer numbers of observer events to which particle events project. The values under this change in meaning were called effective in \cite{Walsh+Knuth:MaxEnt2015} to emphasize that the values were no longer numbers of events or lengths per each and every event on the observer chains.

\section{Newton's Laws of Motion}

Theorem \ref{thm:constancy_of_k} and Lemma \ref{lemma:k_pure} are formal presentations of a fact that was alluded to in \cite{Knuth+Bahreyni:JMP2014}.

\begin{lemma}\label{lemma:k_pure}
The value of $k$ cannot change on an atomic  pure spacetime interval.
\end{lemma}

\textbf{Proof} Theorem \ref{thm:old_thm_3_spacetime} states that there is a particular value of $k$ for a spacetime interval. Since the interval here is an atomic pure spacetime interval, there is no smaller spacetime interval within it such that there could be a different value of $k$ than the one associated with the entire interval. Therefore, $k$ cannot change within the interval.

\begin{theorem}\label{thm:constancy_of_k}
A particle's value of $k$ cannot change between an atomic spacetime interval and a succeeding atomic pure spacetime interval  sharing exactly one event.
\end{theorem}

\textbf{Proof} By condition 0) following Definition 2, we get that atomic intervals 1 and 2 obey summing of interval length as found from the function $ds$: $ds = ds_1 + ds_2$. Applying condition 1) then gives the following. 
\begin{equation}
d \tau = \frac{N_1}{2} + \frac{N_2}{2} = \sqrt{dp_1 dq_1} + \sqrt{dp_2 dq_2} = \sqrt{(dp_1 + dp_2)(dq_1 + dq_2)}
\end{equation} 
Squaring and simplifying gives
\begin{equation}\label{eq:sq_and_sim}
2 \sqrt{dp_1 dq_2 dp_2 dq_1} = dp_1 dq_2 + dp_2 dq_1.
\end{equation}
The restriction of the second interval to the pure status is required to keep its value of $k$ constant by Lemma  \ref{lemma:k_pure}. Since intervals 1 and 2 obey the conditions of Definition 2, equation (\ref{eq:sq_and_sim}) becomes
\begin{equation}
2 \sqrt{\frac{N_1}{2} k_1 \frac{N_2}{2k_2} \frac{N_2}{2} k_2 \frac{N_1}{2k_1}} = \frac{N_1}{2} k_1 \frac{N_2}{2k_2} + \frac{N_2}{2} k_2 \frac{N_1}{2k_1}
\end{equation}
\begin{equation}
2 \frac{N_1}{2} \frac{N_2}{2} = \frac{N_1}{2} \frac{N_2}{2} \bigg[ \frac{k_1}{k_2} + \frac{k_2}{k_1}   \bigg].
\end{equation}
With $A = \frac{k_1}{k_2}$,
\begin{equation}
2 = A + \frac{1}{A}, \qquad \qquad \mathrm{or} \qquad \qquad A^2 - 2A + 1 = 0.
\end{equation}
This has solution $A = 1$, meaning $k_1 = k_2$.

Theorem \ref{thm:constancy_of_k} allows the possibility that the first interval between spacetime events ends in a reception of influence from another particle. As explored in \cite{Walsh+Knuth:MaxEnt2015}, this causes an atomic spacetime interval to be followed by a non-spacetime interval bounded by spacetime events, so that the union of the two intervals is a \textit{mixed}, rather than pure,  spacetime interval. In this process, $k$ changes to an effective value that holds for the union of the two intervals, which can be considered interval 1 here. Thus, Theorem \ref{thm:constancy_of_k} shows that the new value of $k$ along the first interval is the value of $k$ along the second, formally establishing a fact that was used in \cite{Walsh+Knuth:MaxEnt2015}. Note also that a pure spacetime interval is an atomic spacetime interval, so that $k$ does not change between atomic pure spacetime intervals that compose a pure spacetime interval.

Lemma \ref{lemma:k_pure} and Theorem \ref{thm:constancy_of_k} are useful due to the central role of $k$ in characterizing influence sequences. In turn, the influence sequences can be described at spacetime events in terms of motion in spacetime. This leads to Newton's Laws of motion.

\begin{theorem}\label{thm:First_Law}
In the absence of force, the velocity associated with spacetime intervals in 1+1 dimensions cannot change.
\end{theorem}

\textbf{Proof} Equation (\ref{eq:E_and_P}) presents the expression for momentum in the influence network in 1+1 dimensions found in \cite{Knuth:Info-Based:2014}. From Theorem \ref{thm:old_thm_3_spacetime}, this can be rewritten for a spacetime interval as $P = k - \frac{1}{k}$. With force defined as usual as $F \dot{=} \frac{dP}{d \tau}$, the force over a spacetime interval is 
\begin{equation}
F = \frac{dk}{d \tau} \bigg[ 1 + \frac{1}{k^2} \bigg].
\end{equation}
The quantity in brackets is positive, so $F = 0$ implies that $k$ is constant. From \cite{Knuth+Bahreyni:JMP2014}, we have that $v = \frac{k - 1/k}{k + 1/k}$ on spacetime intervals. Thus, constant $k$ implies constant $v$ on spacetime intervals, and so does $F = 0$.

In \cite{Walsh+Knuth:MaxEnt2015}, it was shown that reception of influence can cause a change in $k$ and hence an acceleration and associated force. Recent unpublished results give the conditions under which $k$ changes in the case examined in \cite{Walsh+Knuth:MaxEnt2015}. This force resembles gravity in 1+1 dimensions in that it gives rise to equations of geodesic form, as in general relativity. Future work will involve examining other influence sequences to determine if other forces arise in the influence network.

\begin{theorem}\label{thm:old_thm_5_2nd_Law}
For spacetime intervals, Newton's Second Law is obeyed in 1+1 dimensions: $F = M \frac{d u}{d \tau}$ and $\mathcal{P} \dot{=} \frac{dE}{d \tau} = Fv$, where $u = \frac{dx}{d \tau} = \gamma v$. 
\end{theorem}

\textbf{Proof} Since we are working in lattice units, we have $M = 2$ from Lemma \ref{thm:old_lemma_3-1_mass_is_2}. Then
\begin{equation}
M \frac{du}{d \tau} = 2 \frac{d}{d \tau} \gamma v = 2 \frac{d}{d \tau} \frac{dt}{d \tau} \frac{dx}{d t} = 2 \frac{d}{d \tau} \frac{dx}{d \tau} =  2 \frac{d}{d \tau} \frac{k - \frac{1}{k}}{2},
\end{equation}
where $dx$ and $d \tau$ have been replaced by $\frac{N}{2} \frac{k - 1/k}{2}$ and $\frac{N}{2}$, respectively, as holds in lattice units 
on spacetime intervals by Theorem \ref{thm:old_thm_3_spacetime} and the fact that the interval length on the particle chain is the proper time. Hence,
\begin{equation}
M \frac{du}{d \tau} = \frac{d}{d \tau} \bigg(k - \frac{1}{k}     \bigg) = \frac{d P}{d \tau} = F.
\end{equation}
For the power on spacetime intervals, we have again from (\ref{eq:E_and_P}), Theorem \ref{thm:old_thm_3_spacetime}, and the fact that $d \tau = \frac{N}{2}$ that
\begin{equation}
\frac{dE}{d \tau} = \frac{d}{d \tau} \bigg( k + \frac{1}{k} \bigg) = \frac{d k}{d \tau} \bigg(1 - \frac{1}{k^2}  \bigg) = F \frac{1 - \frac{1}{k^2}}{1 + \frac{1}{k^2}} = F \frac{k - \frac{1}{k}}{k + \frac{1}{k}} = F v.
\end{equation}
This is the correct relativistic relation for power, since $E = \sqrt{P^2 + M^2}$ gives $\frac{dE}{d \tau} = \frac{dP}{d \tau} \frac{P}{E} = F v$.

\begin{figure}
\centering
\includegraphics[width=0.35\linewidth]{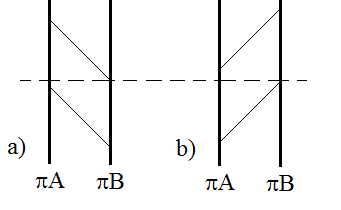}
\caption{These are the Hasse diagrams considered in Theorem \ref{thm:old_thm_6_rates_do_not_change}. Chains $\pi_A$ and $\pi_B$ are particles lying between observers $\P$ and $\Q$, which are not pictured. Each of the two diagrams is the other diagram with the ordering relation reversed, corresponding to reflection through the dotted line. Not depicted are the events of influencing the observers that lie between the illustrated events.}
\label{fig:Momentum_Conservation}
\end{figure}

\begin{theorem}\label{thm:old_thm_6_rates_do_not_change}
Successive receptions by one particle at spacetime events of influence sent from spacetime events of another particle do not change the total rates at which observer events are projected to by particle events in the case that both influences come from the same side.
\end{theorem}

\textbf{Proof} In Figure  \ref{fig:Momentum_Conservation}a, particle $\pi_A$ receives two influences from particle $\pi_B$, which remains on the same side of $\pi_A$ during the process. From \cite{Walsh+Knuth:MaxEnt2015}, $k_A$ on the interval along $\pi_A$ following the first reception changes due to receipt of the second influence from $\pi_B$ as follows.
\begin{equation}\label{eq:change_in_kA}
k_A' = \frac{N_A + 1}{N_A} k_A,
\end{equation}
in which a prime indicates that the quantity applies after receipt of the second influence. Arguing at this point simply from reasonableness, in order for the second reception to occur, the increment in $p$ for the two particles since the first reception must match in the absence of changes due to the reception. Otherwise, the second pair of events on the two particle chains would be at incompatible positions in the order. From \cite{Walsh+Knuth:MaxEnt2015} and the fact that the events are spacetime in nature,
\begin{equation}\label{eq:A_reception_relation}
\frac{N_A+1}{2} k_A = \frac{N_B}{2} k_B.
\end{equation}
Since all events under consideration are assumed to be in the 1+1 dimensional subspace that arises from observers $\P$ and $\Q$, the illustrated events are properly collinear with their projections onto the observers. This means that the ordering relation can be reversed without changing the projections \cite{Knuth+Bahreyni:JMP2014}, as illustrated in Figure  \ref{fig:Momentum_Conservation}b, the reversal being equivalent to reflection of Figure \ref{fig:Momentum_Conservation}a across the dotted line. In the second diagram, $\pi_A$ influences $\pi_B$, meaning that whatever happens to $\pi_B$ due to its influencing $\pi_A$ must be equivalent in some way to receiving influence from $\pi_A$. This gives relations analogous to (\ref{eq:change_in_kA}) and (\ref{eq:A_reception_relation}), again from \cite{Walsh+Knuth:MaxEnt2015}:
\begin{equation}\label{eq:B_conditions}
k_B' = \frac{N_B}{N_B+1} k_B \qquad \qquad \mathrm{and} \qquad \qquad \frac{N_A}{2} \frac{1}{k_A} = \frac{N_B+1}{2} \frac{1}{k_B},
\end{equation} 
the differences between the two pairs of equations being due to the fact that $\pi_A$ receives from its $\Q$ side in Figure \ref{fig:Momentum_Conservation}a and $\pi_B$ from its $\P$ side in Figure \ref{fig:Momentum_Conservation}b. Using (\ref{eq:change_in_kA}) and the first relation of (\ref{eq:B_conditions}), we can now calculate the total rates of influencing the observers after the second reception.
\begin{equation}
r_{pA}' + r_{pB}' = \frac{2}{k_A'} + \frac{2}{k_B'} = 2 \bigg[ \frac{N_A}{N_A + 1} \frac{1}{k_A} + \frac{N_B+1}{N_B} \frac{1}{k_B} \bigg] = 2 \bigg[ \frac{1}{k_A} + \frac{1}{k_B} -  \frac{1}{N_A + 1} \frac{1}{k_A} + \frac{1}{N_B} \frac{1}{k_B} \bigg]
\end{equation}
Applying the condition expressed in (\ref{eq:A_reception_relation}) causes the last two terms to cancel, giving $r_{pA}' + r_{pB}' = \frac{2}{k_A} + \frac{2}{k_B} = r_{pA} + r_{pB}$. Proceeding in the same way with the total rate at which $\Q$ is influenced gives $r_{qA}' + r_{qB}' = r_{qA} + r_{qB}$, the second half of the desired result.

\begin{lemma}\label{thm:old_lemma_6-1_3rd_Law}
Under the conditions of Theorem \ref{thm:old_thm_6_rates_do_not_change}, energy and momentum are conserved.
\end{lemma}

\textbf{Proof} This follows immediately from the definitions of energy and momentum, (\ref{eq:E_and_P}), which depend only on the rates of influencing the observers.

Equation (\ref{eq:A_reception_relation}) and the second relation in (\ref{eq:B_conditions}), as well as the fact that $dp_A' = dp_B'$, which can be seen from Figure \ref{fig:Momentum_Conservation}a, place constraints on reception of influence connecting spacetime events. This will be explored in future work.

\section{Conclusion}

Here, we have classified certain sequences of events in the influence network on the basis of collinearity and the spacetime property, finding that Newton's First and Second Laws of Motion hold in all cases and that Newton's Third Law, in the form of momentum conservation, is obeyed under the conditions of Theorem \ref{thm:old_thm_6_rates_do_not_change}. We expect that Figure \ref{fig:Sequence_Classification}a$'$ will give a substantial hint to the details by which the conservation laws of Lemma \ref{thm:old_lemma_6-1_3rd_Law} arise. Future work will also involve extending these results into 3+1 dimensions, using the work on higher dimensions of one of us (Knuth).

\section{ACKNOWLEDGMENTS}

We wish to thank Newshaw Bahreyni, Ariel Caticha, Seth Chaiken, Philip Goyal, Keith Earle, Oleg Lunin, Anthony Garrett, John Skilling, Kevin Vanslette, Selman Ipek, and Ruth Kastner  for interesting discussions and helpful questions and comments.

\bibliographystyle{aipnum-cp}%
\bibliography{Walsh}%

\begin{thebibliography}{4}%
\makeatletter
\providecommand \@ifxundefined [1]{%
 \@ifx{#1\undefined}
}%
\providecommand \@ifnum [1]{%
 \ifnum #1\expandafter \@firstoftwo
 \else \expandafter \@secondoftwo
 \fi
}%
\providecommand \@ifx [1]{%
 \ifx #1\expandafter \@firstoftwo
 \else \expandafter \@secondoftwo
 \fi
}%
\providecommand \natexlab [1]{#1}%
\providecommand \enquote  [1]{``#1''}%
\providecommand \bibnamefont  [1]{#1}%
\providecommand \bibfnamefont [1]{#1}%
\providecommand \citenamefont [1]{#1}%
\providecommand \href@noop [0]{\@secondoftwo}%
\providecommand \href [0]{\begingroup \@sanitize@url \@href}%
\providecommand \@href[1]{\@@startlink{#1}\@@href}%
\providecommand \@@href[1]{\endgroup#1\@@endlink}%
\providecommand \@sanitize@url [0]{\catcode `\$12\catcode `\&12\catcode
  `\#12\catcode `\^12\catcode `\_12\catcode `\%12\relax}%
\providecommand \@@startlink[1]{}%
\providecommand \@@endlink[0]{}%
\providecommand \url  [0]{\begingroup\@sanitize@url \@url }%
\providecommand \@url [1]{\endgroup\@href {#1}{\urlprefix }}%
\providecommand \urlprefix  [0]{URL }%
\providecommand \Eprint [0]{\href }%
\providecommand \doibase [0]{http://dx.doi.org/}%
\providecommand \selectlanguage [0]{\@gobble}%
\providecommand \bibinfo  [0]{\@secondoftwo}%
\providecommand \bibfield  [0]{\@secondoftwo}%
\providecommand \translation [1]{[#1]}%
\providecommand \BibitemOpen [0]{}%
\providecommand \bibitemStop [0]{}%
\providecommand \bibitemNoStop [0]{.\EOS\space}%
\providecommand \EOS [0]{\spacefactor3000\relax}%
\providecommand \BibitemShut  [1]{\csname bibitem#1\endcsname}%
\let\auto@bib@innerbib\@empty
\bibitem [{\citenamefont {Walsh}\ and\ \citenamefont
  {Knuth}(2016)}]{Walsh+Knuth:MaxEnt2015}%
  \BibitemOpen
  \bibfield  {author} {\bibinfo {author} {\bibfnamefont {J.~L.}\ \bibnamefont
  {Walsh}}\ and\ \bibinfo {author} {\bibfnamefont {K.~H.}\ \bibnamefont
  {Knuth}},\ }\bibfield  {booktitle} {\emph {\bibinfo {booktitle} {Bayesian
  Inference and Maximum Entropy Methods in Science and Engineering, Potsdam,
  New York, USA, 2015}},\ }\href@noop {} {\ \bibinfo {series} {AIP Conf. Proc.}
  (\bibinfo {year} {2016})},\ \bibinfo {note} {(arXiv:1604.08112
  [quant-ph])}\BibitemShut {NoStop}%
\bibitem [{\citenamefont {Knuth}\ and\ \citenamefont
  {Bahreyni}(2014)}]{Knuth+Bahreyni:JMP2014}%
  \BibitemOpen
  \bibfield  {author} {\bibinfo {author} {\bibfnamefont {K.~H.}\ \bibnamefont
  {Knuth}}\ and\ \bibinfo {author} {\bibfnamefont {N.}~\bibnamefont
  {Bahreyni}},\ }\href@noop {} {\bibfield  {journal} {\bibinfo  {journal} {J.
  Math. Phys.}\ }\textbf {\bibinfo {volume} {55}},\ p.\ \bibinfo {pages}
  {112501} (\bibinfo {year} {2014})},\ \bibinfo {note} {(arXiv:1209.0881
  [math-ph])}\BibitemShut {NoStop}%
\bibitem [{\citenamefont {Knuth}(2010)}]{Knuth:infophysics}%
  \BibitemOpen
  \bibfield  {author} {\bibinfo {author} {\bibfnamefont {K.~H.}\ \bibnamefont
  {Knuth}},\ }in\ \href@noop {} {\emph {\bibinfo {booktitle} {Bayesian
  Inference and Maximum Entropy Methods in Science and Engineering, Chamonix,
  France, 2010}}},\ \bibinfo {series and number} {AIP Conf. Proc. 1305},\
  \bibinfo {editor} {edited by\ \bibinfo {editor} {\bibfnamefont
  {P.}~\bibnamefont {Bessiere}}, \bibinfo {editor} {\bibfnamefont {J.-F.}\
  \bibnamefont {Bercher}}, \ and\ \bibinfo {editor} {\bibfnamefont
  {A.}~\bibnamefont {Mohammad-Djafari}}}\ (\bibinfo  {publisher} {AIP},\
  \bibinfo {address} {New York},\ \bibinfo {year} {2010})\ \unskip, pp.\
  \bibinfo {pages} {3--19},\ \bibinfo {note} {(arXiv:1009.5161v1
  [math-ph])}\BibitemShut {NoStop}%
\bibitem [{\citenamefont {Knuth}(2014)}]{Knuth:Info-Based:2014}%
  \BibitemOpen
  \bibfield  {author} {\bibinfo {author} {\bibfnamefont {K.~H.}\ \bibnamefont
  {Knuth}},\ }\href@noop {} {\bibfield  {journal} {\bibinfo  {journal}
  {Contemporary Physics}\ }\textbf {\bibinfo {volume} {55}},\ \unskip\ \bibinfo
  {pages} {12--32} (\bibinfo {year} {2014})}\BibitemShut {NoStop}%
\end{thebibliography}%

\end{document}